\documentclass{ws-p8-50x6-00}
\def\RMP{\em Rev. Mod. Phys.}
\def\EPJC{{\em Eur. Phys. J.} C}
\def\PR{\em Phys. Rep.}
\def\SJNP{\em Sov. J. Nucl. Phys.}
\def\PRC{{\em Phys. Rev.} C}
\def\JCP{\em J. Comput. Phys.}%
\def\htm{\hat{t}}
\def\la{\langle}
\def\ra{\rangle}
\def\kav{\la k_T \ra}
\def\k2av{\la k_T^2 \ra}
\def\knav{\la k_T^2 \ra}
\newcommand{\f}[2]{\frac{#1}{#2}}
\begin{document}

\title{HARD PHOTONS AND NEUTRAL PIONS FROM RHIC}

\author{G\'ABOR PAPP, GEORGE FAI}

\address{Center for Nuclear Research, Department of Physics, 
Kent State University, Kent\\ OH 44242, USA} 

\author{P\'ETER L\'EVAI}

\address{KFKI Research Institute for Particle and Nuclear Physics,
P.O. Box 49,\\
Budapest, 1525, Hungary}

\maketitle

\abstracts{In order to fix the parameters for predictions of hard photon and pion 
production in $Au+Au$ collisions at $\sqrt{s}=200$ GeV, proton-proton and proton-nucleus
data are analyzed in perturbative QCD in the energy range $\sqrt{s} \approx 20-60$ GeV.}
\vspace*{-7mm}
\section{Introduction}

Interest in direct photon production at RHIC\cite{phenix,star} 
is motivated by the long mean free path of photons. 
Hard photon production can be calculated in the pQCD-improved parton model. 
Experiments need to separate the direct photons 
from the $\pi^0$ (and $\omega^0$) two-photon background. 
It is estimated that the direct photon yield has to be at least 10\% 
of the $\pi^0$ yield to be measurable. We have embarked on the 
following project: (i) fix the pQCD description of $\gamma$ and $\pi^0$
production in $pp$ collisions; (ii) address nuclear effects in $pA$
collisions; (iii) calculate direct $\gamma$ and $\pi^0$
production in $AA$ collisions at $\sqrt{s}=200$ GeV.
Our first results, which pertain to (i) and (ii), are summarized in Ref. 3.
 
Parton cross sections are calculable in pQCD at high energy 
to leading order (LO) or next-to-leading-order (NLO).\cite{Field,NLOAur} 
The parton distribution functions (PDFs) and fragmentation functions (FFs),
however, need to be fitted to data. In recent NLO calculations the various scales 
are optimized.~\cite{OptimQres} Alternatively,
an additional non-perturbative parameter, the width of the {\it intrinsic transverse
momentum} ($k_T$) distribution of the partons is introduced.~\cite{Huston95} 
We choose the latter method and use $k_T$ phenomenologically in $pp$ collisions,
expecting that its importance will decrease in higher orders of pQCD.

\nopagebreak
\section{Proton-proton collisions}\label{sec:pp}

In the lowest-order pQCD-improved parton model,
direct pion production can be described in $pp$ collisions by
\be
\label{fullpi}
  E_{\pi}\f{d\sigma_\pi^{pp}}{d^3p} =
        \sum_{abcd}\!  \int\!dx_{1,2}\ f_{1}(x_1,Q^2)\
        f_{2}(x_2,Q^2)\  K  \f{d\sigma}{d\htm}(ab\to c d)\,
   \frac{D(z_c,{\widehat Q}^2)}{\pi z_c} \ \ \  , 
\ee
where  $f_{1}(x,Q^2)$ and  $f_{2}(x,Q^2)$  are the PDFs of partons $a$ and $b$, and
$\sigma$ is the LO cross section of the appropriate
partonic subprocess. The K-factor accounts for higher order corrections.~\cite{Owens87} 
Comparing LO and NLO calculations a nearly constant value, $K\approx 2$, 
is obtained as a good approximation of the higher order contributions 
in the $p_T$ region of interest.~\cite{Wong98} In eq.(\ref{fullpi}) 
$D(z_c,{\widehat Q}^2)$ is the FF of the pion,
with ${\widehat Q} = p_T /z_c$, where 
$z_c$ is the momentum fraction of the final hadron.
We use NLO parameterizations of the
PDFs\cite{MRST98} and FFs~\cite{BKK} with fixed scales.  
Direct $\gamma$ production is described similarly.~\cite{Field}

\begin{figure}[b]
\centerline{\epsfysize=2in\epsfbox{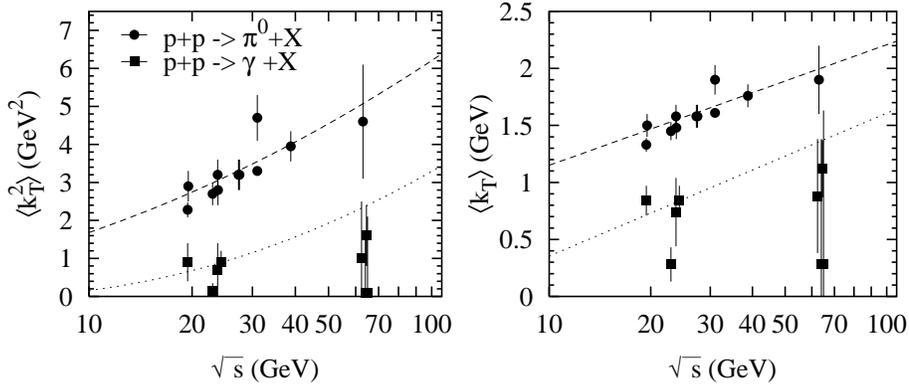}}
\caption{Best fits for the widths of the intrinsic transverse momentum 
distributions in $\gamma$ and $\pi^0$ production data in $pp$
collisions as a function of cm energy. See text for the lines.
\label{fig:kTav}}
\end{figure}

To incorporate the intrinsic $k_T$, each integral is extended to $k_T$-space,
$dx \ f(x,Q^2) \rightarrow dx 
\ d^2\!k_T\ g({\vec k}_T) \  f(x,Q^2)$.~\cite{Wong98,Wang9798}
We approximate $g({\vec k}_T)$ as
$g({\vec k}_T) = \exp(-k_T^2/\langle k_T^2 \rangle)/
{\pi \langle k_T^2 \rangle}$.
Here $\langle k_T^2 \rangle$ is the 2-dimensional width of the $k_T$
distribution, related to the average $k_T$ of one parton
as $\langle k_T^2 \rangle = 4 \langle k_T \rangle^2 /\pi$.

We applied this model to data from
$pp\rightarrow \pi^0  X$  and $pp\rightarrow \gamma X$.\cite{Cronin,pi,gamma}
The calculations were corrected for the finite rapidity windows of 
the data. The Monte-Carlo integrals were carried out by the
standard VEGAS routine.~\cite{VEGAS}
We fitted the data minimizing  $\Delta^2=\sum (Data-Theory)^2/Theory^2$
in the midpoints of the data. Fig. 1. shows the obtained fit values for
$\knav$ and a calculated $\kav$. The error bars display a 
$\Delta^2= \Delta^2_{min}\pm 0.1$ uncertainty in the fit procedure.
The lines guide the eye and indicate a linear increase to a common value 
of $\kav \approx 3.5$ GeV at $\sqrt{s}=1800$ GeV.\cite{Ziel98} 
  
These values of $\kav$ provide a satisfactory description of the data
up to energies of $\sqrt{s} \approx 65$ GeV.
The data/theory ratios cluster around unity, with no systematic trend 
in the deviations (different experiments show different slopes).\cite{plf99,plf99;a}
The maximum deviations are on the order of $\approx 50\%$.   

To estimate the $p_T$ value where the direct photon production cross section 
surpasses 10\% of the pion cross section in a $pp$ collision at
$\sqrt{s}=200$ GeV, we use the $\sqrt{s}$ dependence of $\la k_T \ra$
from Fig.~1, and obtain the results displayed in Fig.~2. 
The threshold  is reached in this approximation at $p_T \approx 6.5$ GeV.

\begin{figure}[h]
\centerline{\epsfxsize=2.5in\epsfbox{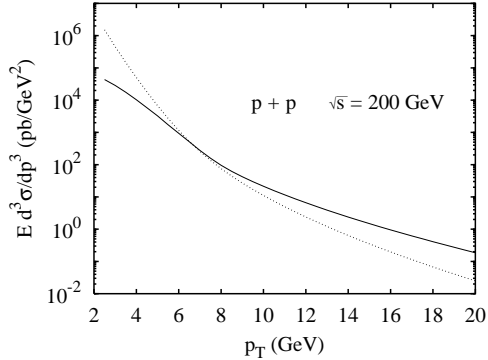}}
\caption{Extrapolated $\gamma$ (full line) and $\pi^0 (\times 0.1)$ 
(dashed) production cross sections as
functions of transverse momentum in $pp$ collisions at $\sqrt{s}=200$ GeV.
\label{fig:pp200}}
\end{figure}
 
\vspace*{-4mm}
      
\section{Proton-nucleus collisions}\label{sec:pA}

Having isolated the $\knav$ already present in 
$\pi^0$ and $\gamma$ production in $pp$ collisions at the
present level of calculation, we turn to the nuclear enhancement
observed\cite{Cronin} in $pA$ collisions.
In minimum-biased $pA$ collisions the pQCD description 
of the inclusive pion cross section is based on
\begin{eqnarray}
\label{fullpipA}
 && E_{\pi}\f{d\sigma_\pi^{pA}}{d^3p} =\!
        \sum_{abcd}\!  \int\!\! d^2\!b\, t_A(b)\!
        \int\!\!dx_{1,2} d^2k_{T_{1,2}}\ 
        g_1(\vec{k}_{T_1},b) 
	\nonumber \\
        && \ \ g_2(\vec{k}_{T_2}) f_{1}(x_1,Q^2) f_{2}(x_2,Q^2)\ 
           K  \f{d\sigma}{d\htm}
   \frac{D(z_c,{\widehat Q}^2)}{\pi z_c} \,.
\end{eqnarray}
Here $b$ is the impact parameter and
$t_A(b)$ is the nuclear thickness function normalized as
$\int d^2 b \ t_A(b) = A$. For simplicity, we  use 
a sharp sphere nucleus with 
$t_A(b) = 2 \rho_0 \sqrt{R_A^2-b^2} $, where $R_A=1.14 A^{1/3}$
and $\rho_0=0.16$ fm$^{-3}$. 

\begin{figure}[b]
\centerline{\epsfxsize=2.3in\epsffile{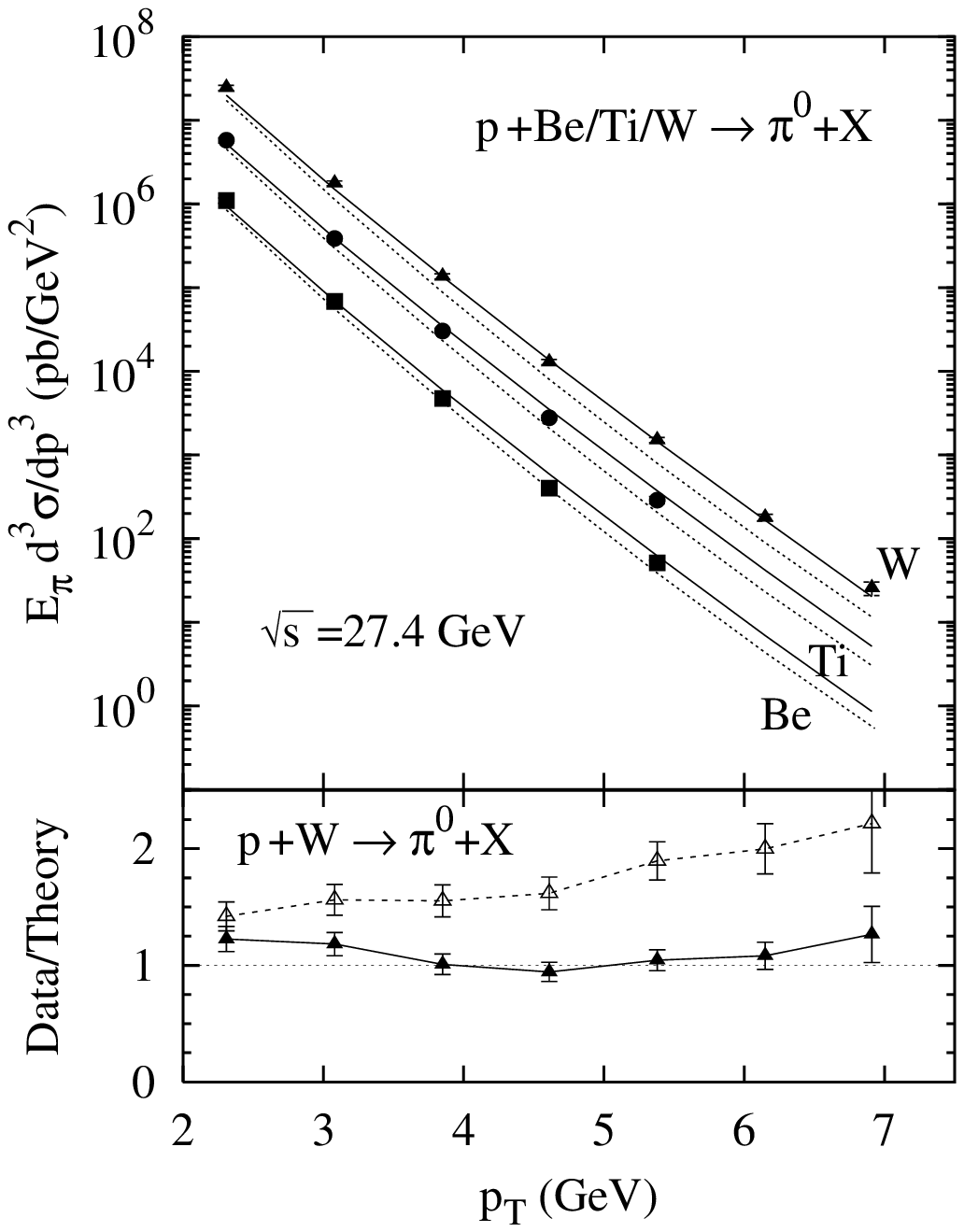}
	\hfill  \epsfxsize=2.3in\epsffile{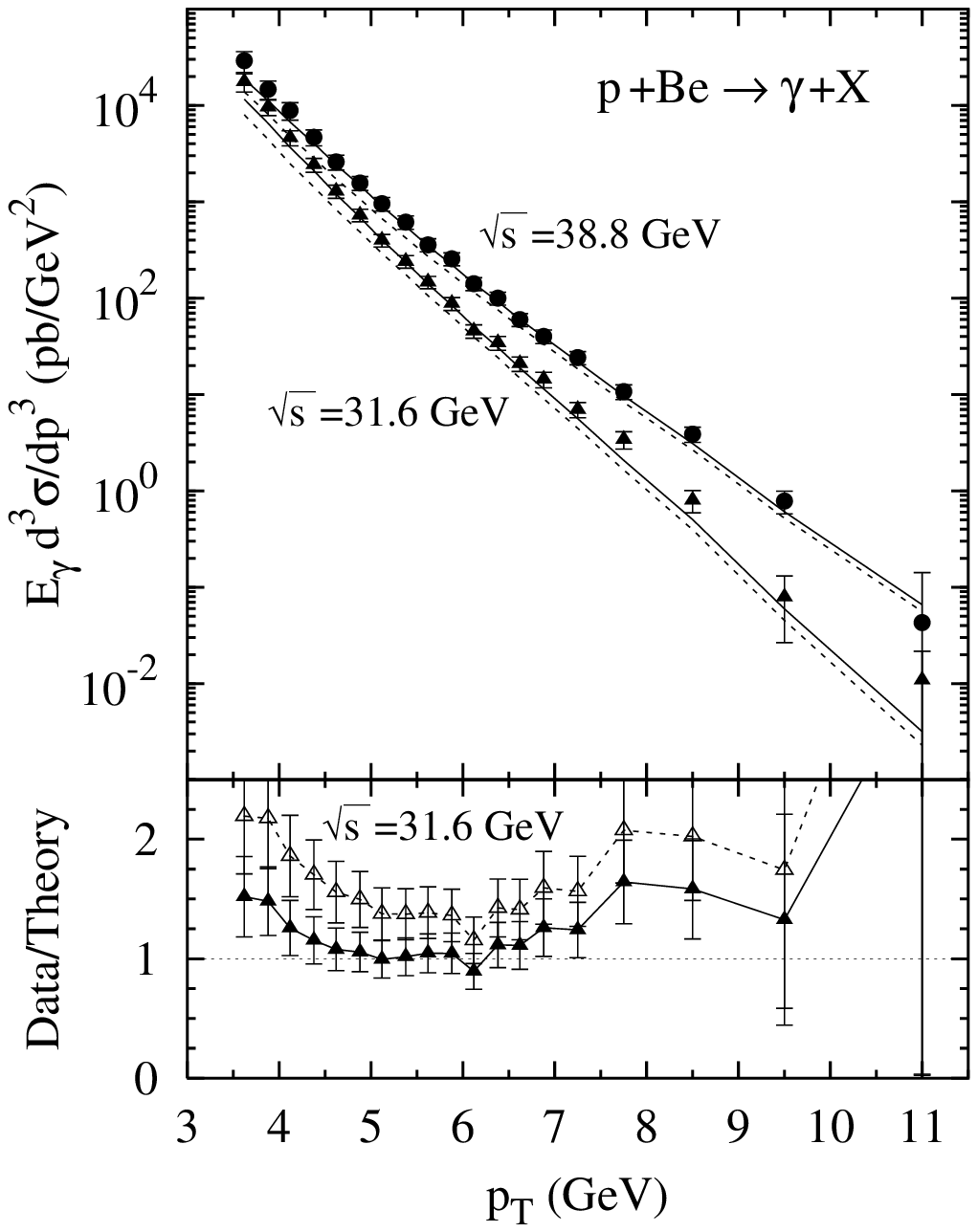}}
\caption[]{{\it Left}: Cross section per nucleon in 
$pA \rightarrow \pi^0 X$ reactions ($A=Be,Ti,W$). We show 
$C^{sat}=1.2$ GeV$^2$ (full lines) and
$C^{sat}=0$ (dashed lines). Lower panel:
data/theory on a linear scale for the $pW$ collision.
{\it Right}: Cross section per nucleon in the 
$pBe \rightarrow \gamma X$ reaction 
at two energies.\cite{E706}
Solid lines indicate $C^{sat}=1.2$ GeV$^2$, dashed lines mean $C^{sat}=0$.
Lower panel shows data/theory for $\sqrt{s}=31.6$ GeV.%
}
\label{fig:pApi}
\end{figure}

The standard physical explanation of the nuclear enhancement
(Cronin effect) is that the proton traveling
through the nucleus gains extra transverse momentum due to 
random soft collisions and the partons enter the final hard
process with this extra  $k_T$.\cite{Wang9798} 
We write  the width of the transverse momentum distribution of the partons
in the incoming proton as
\begin{equation}
\label{ktbroadpA}
\knav_{pA} = \knav_{pp} + C \cdot h_{pA}(b) \ . 
\end{equation}
Here $h_{pA}(b)$ is the number of {\it effective} 
nucleon-nucleon collisions at impact parameter $b$ 
imparting an average transverse momentum squared $C$. 
Naively all possible soft interactions are included,
but such a picture leads to a target-dependent $C$.\cite{plf99}
A more satisfactory description is obtained with a ``saturated''
$h_{pA}$, where it takes
at most one semi-hard ($Q^2 \sim 1$ GeV$^2$) collision
for the incoming proton to loose coherence, 
resulting in an increase of the width of its $k_T$ distribution.
This is approximated by a smoothed step
function with a maximum value of unity.
The saturated Cronin factor is denoted by  $ C^{sat}$.

As displayed in the left of Fig. 3, 
$C^{sat}=1.2$ GeV$^2$ gives a good fit of $p A \rightarrow \pi^0 X$
data with $A=Be$, $Ti$ and $W$.\cite{Cronin}
The right panel of Fig. 3 illustrates how
$C^{sat}=1.2$ GeV$^2$ describes $\gamma$ production
at $\sqrt{s} \approx 30-40$ GeV.

We interpret $C^{sat}$ as the square of the typical transverse
momentum imparted in {\it one} semi-hard collision prior to the hard
scattering. This picture and the energy dependence of $C^{sat}$ need 
to be tested as functions of $A$ at different energies, and in particular
at $\sqrt{s}=200$ GeV, for RHIC $AA$ predictions.

\vspace*{-3mm}
\section*{Acknowledgments}
\vspace*{-1mm}
Supported by DOE, DE-FG02-86ER40251, and by OTKA, F019689.

\end{document}